\documentclass{article}
\usepackage{spconf,amsmath,graphicx}

\title{Can DNNs learn to lipread full sentences?}
\name{George Sterpu, Christian Saam, Naomi Harte}
\address{Sigmedia, ADAPT Centre, School of Engineering, Trinity College Dublin, Ireland\thanks{Supported by the ADAPT Centre for Digital Content Technology, which is funded under the SFI Research Centres Programme (Grant 13/RC/2106) and is co-funded under the European Regional Development Fund.}}
\begin{document}
%
\maketitle
\begin{abstract}
Finding visual features and suitable models for lipreading tasks that are more complex than a well-constrained vocabulary has proven challenging. This paper explores state-of-the-art Deep Neural Network architectures for lipreading based on a Sequence to Sequence Recurrent Neural Network. We report results for both hand-crafted and 2D/3D Convolutional Neural Network visual front-ends, online monotonic attention, and a joint Connectionist Temporal Classification-Sequence-to-Sequence loss. The system is evaluated on the publicly available TCD-TIMIT dataset, with 59 speakers and a vocabulary of over 6000 words. Results show a major improvement on a Hidden Markov Model framework. A fuller analysis of performance across visemes demonstrates that the network is not only learning the language model, but actually learning to lipread.
\end{abstract}
\section{Introduction}
\label{sec:intro}

Automatic lipreading of continuous and large vocabulary speech is a promising technology with many applications, recovering the information in speech from a different modality than the acoustic one. The traditional approaches have largely followed early approaches in speech recognition, using handcrafted features and Hidden Markov Models (HMM). These have been so far unsuccessful at modelling the complex patterns of visual speech \cite{tcdtimit, sterpu2017, uea_interspeech2017, uea_bmvc, review_pota}, and several research problems, such as finding good representations, remain open in the lipreading community. 

The Sequence to Sequence Recurrent Neural Network (Seq2seq RNN) architecture has seen a surge in popularity since it was first introduced in \cite{seq2seq_paper} for machine translation. It has an elegant formulation, makes minimal assumptions about the modelled sequences, requires less domain knowledge and has obtained competitive results on many benchmarks. Together with the Connectionist Temporal Classification (CTC) method \cite{graves_ctc}, these represent the main end-to-end trainable approaches for transcribing temporal patterns. In this work we prefer Seq2seq for its additional property of implicitly learning a language model, as CTC performance is limited by the conditional independence of its predictions \cite{graves_ctc}.

Several recent advancements in machine learning have not been explored by the lipreading community. These include the monotonic attention \cite{monotonic_attn} and the joint CTC-Sequence loss \cite{jointctc}. In addition, several successful applications in Automatic Speech Recognition (ASR) \cite{chorowski2014, chorowski2015, bahdanau2016, las}, using both medium-sized (TIMIT, WSJ) and large (Google Speech Commands) datasets, give us some useful insights from a different, though correlated modality. Yet experience has shown that techniques successful for audio-only speech recognition don't automatically translate well to a lip-reading task \cite{tcdtimit, sterpu2017, uea_interspeech2017, uea_bmvc, review_pota}. Thus our contribution is an exploration of state of the art Seq2seq techniques within the domain of lipreading, to determine what approaches hold the greatest potential in this domain and identify where further challenges remain.

There are, to our best knowledge, only two papers in the literature to date that address the problem of lipreading at the sub-word level using DNNs. The first one \cite{lipnet} uses a spatio-temporal Convolutional Neural Network (CNN) and the CTC loss in order to produce a phonetic transcription of the input sentence. However, the model was applied on a low perplexity dataset, GRID \cite{gridset} where it can be argued that the model can heavily rely on the predictable structure of the sentences. In addition, the CTC loss has its own shortcomings due to the independence assumption for the predicted labels. We address this by testing the algorithms on TCD-TIMIT \cite{tcdtimit}, a dataset of phonetically-balanced sentences and a vocabulary of approx. 6000 words.
The second paper \cite{chung_cvpr_2017} makes use of a spatial only CNN, and applies the Seq2seq network to produce sentence transcriptions at the character level. The dataset used for evaluation, LRS, is larger than TCD-TIMIT but not public, and has recently been superseded by a more challenging, public version, MV-LRS. Our work differs from these two by making predictions at the viseme level, which is a unit choice that avoids ambiguities. In this way, the language model does not have to be well trained in advance, as in \cite{chung_cvpr_2017}. In addition, we explore a wider range of architectures, such as both hand-crafted and 2D/3D CNN visual front-ends, online monotonic attention and a joint CTC-Seq2seq loss. The paper is organised as follows: In Section~\ref{sec:arch} we describe the general network architecture. Section~\ref{sec::eval} presents our experiments, and we discuss our findings in Section~\ref{sec:discuss}.

\section{Model Architecture}
\label{sec:arch}

Our lipreading pipeline has a video processing front-end and a Seq2seq RNN, learning from variable-length videos and producing variable-length transcriptions at the viseme level. The system is trained end-to-end.

\subsection{Visual front-end}

The visual front-end involves segmenting the lip region from a visual stream and computing a feature vector for each frame. We consider both handcrafted features and learnt CNN-based visual representations. At this stage, we can also take advantage of the temporal dimension by appending derivatives or by using 3D convolution kernels.

\subsection{Sequence modelling}

Next, the extracted visual features are fed to a Seq2Seq model, which consists of two RNNs termed as the \emph{encoder} and the \emph{decoder}. With each input timestep, the encoder updates its internal state and produces one output. We collect all the outputs in a \emph{memory} and retain only the final state, known as a \emph{thought vector} that summarises the input sentence. The decoder is initialised from the thought vector and starts producing output symbols from a designated start-of-sentence token until it finally produces an end-of-sentence symbol. As the temporal dimension is warped onto the one-dimensional thought vector, the decoder is allowed to peek into the memory and soft-select the vectors that are correlated with its current internal state. This mechanism is known as \emph{attention}, and the soft-selection temporal pattern is called \emph{alignment}.

With speech signals, enforcing this alignment to be monotonic with respect to the encoded inputs may alleviate the problem of attending to the repetitions of a word in the same sentence. The impact may be more significant for visemes, where the number of classes is typically much lower than for phonemes or characters. In addition, scanning only a past history of the memory enables the on-line application of the lipreading system, further reducing the time complexity. We consider the implementation of \cite{monotonic_attn}, which was shown to outperform related strategies with a minimal loss in accuracy over the softmax attention baseline.

\subsection{Training and decoding}

In the training stage, the entire transcription is available to the decoder. The embedding of the ground-truth symbol gets fed at every time step, but from time to time we replace it with the previously decoded symbol in order to increase the robustness of the network to recover from mistakes. This training process implies that the predicted output transcription has an identical length with the ground-truth transcription, thus a cross-entropy loss function can be applied.  In the evaluation stage, the ground-truth transcription cannot be used, and the decoder is likely to produce a transcription of a different length. Hence, we evaluate the quality of the prediction by computing the Levenshtein edit distance with respect to the ground truth.

Combining the Seq2Seq cross entropy loss with the CTC loss could lead to several benefits. First, the CTC loss forces the encoder to better focus on the input signal, as it tends to become "lazy" due to the power of the implicitly learnt language model on the decoding side. In addition, the encoder should now learn representations that are more closely related to the class labels, as the CTC first predicts a class for each frame, and only later it merges the repeated symbols. 

\section{Evaluation}
\label{sec::eval}

\subsection{Dataset}

We performed our experiments on TCD-TIMIT \cite{tcdtimit}, a publicly available audio-visual dataset with 59 subjects, each reciting 98 phonetically balanced sentences from a vocabulary of 6000 words, totalling around 8 hours of recordings. Sentences vary from 10 to 65 visemes in length. Evaluation was done on the speaker-dependent protocol of \cite{tcdtimit}, choosing 67 and 31 sentences from each speaker for train and test respectively. The dialect-dependent sentences (name begins with \emph{SA}) were removed. As in the original TIMIT database, these two sentences were common across all speakers. Early results demonstrated that the models quickly learned the structure of these sentences, giving misleading high performance.  Our labels are the same viseme level transcriptions as in \cite{tcdtimit}, which were obtained from a phonetic transcription by mapping phonemes into 12 viseme clusters.

\subsection{Setup}

\textbf{Visual features.}
As the lip region coordinates were already provided in \cite{tcdtimit, sterpu2017}, we used them to crop this region from the video frames, downsampled to 36x36 pixels and converted it to grey scale as a preprocessing step. We first considered handcrafted features and kept 44 low frequency coefficients of the lip region 2D DCT transform, plus their first two derivatives, as in \cite{tcdtimit, sterpu2017}. To check the impact of the implicitly learnt language model alone, we also present the results in the absence of a visual stream by replacing the features with zeros. 

Next, we tested multiple CNN architectures on the previously cropped region, additionally using a 36x36 RGB version and a 64x64 grey one to check the benefits of color and a larger window size. Our 2D CNNs have 4 layers with 16, 32, 64 and 128 feature detectors respectively, a small 3x3 convolution kernel and rectified linear activations. After the first layer, our convolutions use a stride of 2 to reduce the dimensionality. The activations of the last layer are flattened and fully connected to a new layer of 128 units, producing our frame-wise feature vectors. The 3D CNN is of the same structure, differing only in the use of a 3x3x3 convolution kernel.


\noindent \textbf{Encoder-decoder RNN.} For our Seq2Seq model we start with two unidirectional recurrent layers of 128 Long Short-term Memory (LSTM) cells each, for both the encoder and the decoder. The one layer version was not performing well and we do not report these results. However, we test a one layer bidirectional LSTM (BiLSTM) version, processing the sentence both in the forward and backward directions, while maintaining the same number of parameters. Decoding was performed using a beam search strategy of width equal to 4.

\noindent \textbf{Attention.} Our default attention mechanism was the Luong \cite{luong_attn} version with the energy term scaled, and we obtained significantly worse results with the more popular Bahdanau attention style \cite{bahdanau_attn}. We also tested the online monotonic attention strategy of \cite{monotonic_attn}. To make it work, we found it was essential to turn off the pre-sigmoid noise and set the scalar bias to a negative value.

\noindent \textbf{Joint CTC-Seq2seq loss.} As the Seq2seq language model exhibits a strong early influence in training, we try to add a CTC loss over the encoder's outputs, inserting a softmax layer over the vocabulary size plus 1, and training jointly with the cross-entropy loss on the decoder side. Since \cite{jointctc} obtained the best results for a mixing coefficient of 0.2 for the CTC loss, we only consider this case here.

\subsection{Practical aspects}

\noindent \textbf{Input pipeline.} We noticed a consistent improvement when randomly shuffling the train files with each dataset iteration. Grouping sentences of similar lengths together, a concept known as bucketing, leads to a smaller zero padding of batches, noticeably reducing the RNN processing time. Our bucket width was 15 frames, or approximately 0.5 seconds.

\noindent \textbf{Regularisation.} We generally obtained good results with dropout applied to the recurrent cells \cite{dropout_rnn}, keeping the inputs, the states and the outputs with a probability of 0.9. For the best results with the CNN architectures, we interleaved dropout layers with a rate of 50\% between convolutions. 
We also applied L2-norm regularisation on the recurrent and the convolutional weights, scaled by 0.0001 and 0.01 respectively. We enable gradient clipping to a maximum norm of 10.0 and we also clip the LSTM cells between -10.0 and 10.0.

\begin{table}[]
\centering
\caption{Lipreading accuracy on TCD-TIMIT. The right column shows the number of iterations needed to reach convergence (or \emph{nc} for \emph{no convergence}).}
\label{tab:results}
\begin{tabular}{l|l|c}
\hline

\textbf{\hfill Feature \hfill} & \textbf{\hfill Accuracy \hfill } & \textbf{Iters} \\ \hline
\textbf{A}. DCT + HMM baseline \cite{sterpu2017} & 31.59 \%  &  -\\ \hline
\textbf{B}. AAM + HMM baseline \cite{sterpu2017} & 25.28 \%  & -\\ \hline
\textbf{C}. Eigenlips + DNN-HMM \cite{uea_bmvc} & 46.61 \%   & - \\ \hline \hline

\textbf{D}. zeros + LSTMs & 45.87 \%   & 160 \\ \hline \hline

\textbf{E}. DCT + LSTMs & 61.52 \%   & 250\\ \hline
\textbf{F}. DCT + BiLSTMs & 60.72 \%   & 180\\ \hline
\textbf{G}. \textbf{E} w/o attention & 48.29 \%   & 270\\ \hline
\textbf{H}. \textbf{E} w/ monotonic attention & 61.58 \% & 170\\ \hline
\textbf{I}. DCT + joint CTC-Seq2seq & 61.18 \%   & 180 \\ \hline \hline

\textbf{J}. 2D CNN + LSTMs &  & \emph{nc} \\ \hline
\textbf{K}. 2D CNN + BiLSTMs & \textbf{66.27 \%} & 400 \\ \hline
\textbf{L}. \textbf{J} on RGB + joint CTC-Seq2seq & 66.20\%   & 150\\ \hline
\textbf{M}. \textbf{J} on 64x64 + joint CTC-Seq2seq &  & \emph{nc} \\ \hline
\textbf{N}. Gray 3D CNN + LSTMs &  & \emph{nc} \\ \hline
\textbf{O}. 2D CNN + joint CTC-Seq2seq & 64.61\%   & 260\\ \hline \hline

\end{tabular}

\end{table}

\section{Discussion and conclusion}
\label{sec:discuss}

The results of our study are shown in Table~\ref{tab:results}. We first observe a massive improvement over the HMM baseline. However, a large part is owed to the implicitly learnt RNN-based language model, as hypothesised in \cite{bahdanau2016} and revealed by system \textbf{D}. In comparison, a bi-gram language model only increased the accuracy of the HMM system \textbf{A} up to 35\% \cite{sterpu2017} on the same dataset, using the same DCT features. Looking at the  predictions, we note that the model quickly learns to output only two visemes in an interleaved pattern, surrounded by the silence visemes delimiting the start and the end of each sentence. These correspond to the \emph{Lips relaxed, narrow opening} and \emph{Tongue up or down} classes, and together they account for 52.56\% of the occurrences in TCD-TIMIT scripts. Since the scripts were phonetically balanced, the viseme distribution only reflects a natural speech pattern.

We identified this matter in all our experiments, typically taking at least 100 iterations before the predictions start to look diverse. This suggests that the language model might slow down training convergence, as the system will learn the patterns from the input signal more slowly.


The use of DCT features with a Seq2seq model led to a substantial improvement over the state of the art on the TCD-TIMIT dataset \cite{uea_bmvc}. There is a noticeable boost in convergence speed from unidirectional to bidirectional LSTMs, yet it does not always translate into higher accuracy, as demonstrated by \textbf{E} and \textbf{F}. This could be explained by the fact that two single-layer networks are less powerful than a single two-layer variant. We tried another variant of two-layered bidirectional LSTM which did not improve the performance.

As noted by \cite{chung_cvpr_2017}, the attention-less system \textbf{G} could not learn meaningful patterns from the input, predicting a similar transcription for most sentences. This could imply that either the temporal information vanishes during encoding, or the decoding process relies heavily on the language model. The attention-based system \textbf{E} alleviates these aspects, obtaining an absolute 13.23\% improvement over this variant.

\begin{figure}

\begin{minipage}{.48\linewidth}
  \centering
  \centerline{\includegraphics[height=5.4cm]{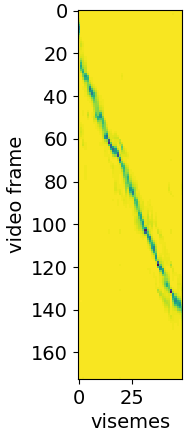}}
  \centerline{System \textbf{K}}\medskip
\end{minipage}
\hfill
\begin{minipage}{0.48\linewidth}
  \centering
  \centerline{\includegraphics[height=5.4cm]{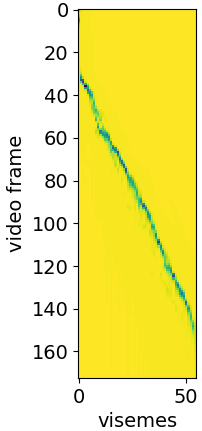}}
  \centerline{System \textbf{H}}\medskip
\end{minipage}
\caption{Typical alignments learnt by our systems}
\label{fig:res}
\end{figure}

Replacing the Luong-style softmax attention with the monotonic attention of \cite{monotonic_attn} maintains the performance at the same level. This is also demonstrated by the alignments in Figure~\ref{fig:res}, where the softmax attention learns to align monotonically, producing a sharp peak in the weight distribution. Consequently, the enforced monotonic attention would represent a suitable choice for lipreading, further reducing the time complexity and enabling online decoding. Cited as a possible extension in \cite{chung_cvpr_2017}, our benchmark shows the first successful application of online and monotonic attention to lipreading.

The use of 2D-CNN features led to an additional $\approx$5\% absolute improvement over the best performing DCT-based system, as is the case with system \textbf{K}. In this case, using BiLSTMs was crucial to prevent the system from getting stuck in a local mininum, as in \textbf{J}. However, our experiments on images of increased resolution (64x64) and with 3D convolutions did not reach convergence, showing the limits of a shallow CNN architecture. 



The use of the joint CTC-Seq2seq loss function significantly accelerates the training process. However, in our case, the test set accuracy was lower than for the cross-entropy loss function alone. The impact of the CTC loss may be twofold. It enforces a frame-wise classification on the encoder's outputs, which could lead to better gradients for the CNN layers. This is demonstrated by the performance achieved with systems \textbf{L} and \textbf{O}, which could not converge without the additional CTC loss. On the other hand, the two loss functions could have competing requirements for the state representation, and a proper weighting may be vital for optimal performance, as shown in \cite{jointctc}.

On the alignments produced by the decoder we could observe that they tend to get fuzzy towards the end of the sentence, sometimes resembling to a river delta. This suggests that the thought vector is quite good at summarising the recent past, and the attention is only needed to boost the decoding of early events. We hypothesise that a different assignment of the thought vector and attention duties, where the first encodes a rather short history and the latter attends to more distant key frames, could enhance the overall performance.


\begin{table}[t]
\centering
\footnotesize{} 
\caption{Viseme accuracy of the best DNN system \textbf{K} and relative change from HMM baseline (\textbf{A}). Visemes sorted by decreasing visibility.}
\label{viseme_inc}
\begin{tabular}{p{2.0cm}|p{2cm}|p{1.4cm}|p{1.5cm}}
\hline
\textbf{Viseme} & \textbf{TIMIT Phonemes} & \textbf{Accuracy K [\%]} & \textbf{$\Delta$ Accuracy K - A [\%]} \\ \hline
Lips to teeth                                         & /f/ /v/                                                          & 85.6
                   & 21.25               \\ \hline
Lips puckered                                         & /er/ /ow/ /r/ /q/ /w/ /uh/ /uw/ /axr/ /ux/                        & 83.4
                   & 50.81               \\ \hline
Lips together                                         & /b/ /p/ /m/ /em/                                                 & 94.8
                   & 30.40               \\ \hline
Lips relaxed moderate opening to lips narrow-puckered & /aw/                                                          & 45.7
                   & 25.90               \\ \hline
Tongue between teeth                                  & /dh/ /th/                                                        & 58.4
                   & 27.79               \\ \hline
Lips forward                                          & /ch/ /jh/ /sh/ /zh/                                              & 65.4
                   & 18.26               \\ \hline
Lips rounded                                          & /oy/ /ao/                                                        & 31.6
                   & -8.41               \\ \hline
Teeth Approximated                                    & /s/ /z/                                                          & 81.6
                   & 52.24               \\ \hline
Lips relaxed narrow opening                           & /aa/ /ae/ /ah/ /ay/ /ey/ /ih/ /iy/ /y/ /eh/ /ax-h/ /ax/ /ix/      & 95.6
                   & 73.50               \\ \hline
Tongue up or down                                     & /d/ /l/ /n/ /t/ /el/ /nx/ /en/ /dx/                               & 84.8
                  & 56.17               \\ \hline
Tongue back                                           & /g/ /k/ /ng/ /eng/                                               & 63.2
                  & 24.41               \\ \hline
Silence                                               & /sil/ /pcl/ /tcl/ /kcl/ /bcl/ /dcl/ /gcl/ /h\#/ /\#h/ /pau/ /epi/ & 93.6
                   & 0.21                \\ \hline
\end{tabular}

\end{table}

We have compared the viseme confusion matrices of systems \textbf{A}, the DCT + HMM baseline, and \textbf{K}, our top performing DNN-based lipreading system. Table \ref{viseme_inc} shows the relative performance increase across the viseme classes for these two systems. The table also shows the TIMIT phonemes mapped to each viseme class and their visibility, or ease of observation for a human. The improvement from \textbf{A} to \textbf{K} is ubiquitous with the exception of a single viseme corresponding to the \emph{Lips rounded} shape. This viseme is most frequently confused with the \emph{Lips relaxed narrow opening} viseme, suggesting that it is difficult even for the CNN to learn features that disambiguate them. Lower improvements are seen for \emph{Lips forward} and \emph{Tongue back}. The frontal view used as input does not capture any depth information, however the database includes a second view at 30$^{\circ}$ which could be useful for such visemes.

Overall, the Seq2seq model greatly outperforms HMM and hybrid DNN-HMM systems even without CNN-based feature extraction. The fully neural architectures achieved the highest accuracies in our experiments. Additionally, the use of the joint loss function boosted the training convergence and enabled learning visual features on higher dimensional inputs. Lastly, we demonstrate the efficiency of online monotonic attention on this task, a necessary step towards online decoding.

\section{Acknowledgements}
We are grateful to Eugene Brevdo, Marco Forte, Oriol Vinyals and Li Deng for their helpful comments and suggestions. 


\bibliographystyle{IEEEbib}
\bibliography{mybib}

\end{document}